# On the abundance of extraterrestrial life after the Kepler mission

Amri Wandel, Racah Inst. of Physics, The Hebrew University of Jerusalem
E-mail: amri@huji.ac.il

## Abstract

The data recently accumulated by the Kepler mission have demonstrated that small planets are quite common and that a significant fraction of all stars may have an Earth-like planet within their Habitable Zone. These results are combined with a Drake-equation formalism to derive the space density of biotic planets as a function of the relatively modest uncertainty in the astronomical data and of the (yet unknown) probability for the evolution of biotic life, $F_b$. I suggest that $F_b$ may be estimated by future spectral observations of exoplanet biomarkers. If $F_b$ is in the range 0.001 -- 1 then a biotic planet may be expected within 10 -- 100 light years from Earth. Extending the biotic results to advanced life I derive expressions for the distance to putative civilizations in terms of two additional Drake parameters - the probability for evolution of a civilization, $F_c$, and its average longevity. For instance, assuming optimistic probability values ($F_b \sim F_c \sim 1$) and a broadcasting longevity of a few thousand years, the likely distance to the nearest civilizations detectable by SETI is of the order of a few thousand light years. The probability of detecting intelligent signals with present and future radio telescopes is calculated as a function of the Drake parameters. Finally, I describe how the detection of intelligent signals would constrain the Drake parameters.

Keywords: Kepler mission – exoplanets – biotic planets – SETI – Drake equation

## Introduction

An important yet until recently poorly known factor, required for estimating the abundance of extraterrestrial life is the fraction of stars with planets, in particular Earth-like planets within the Habitable Zone. The recent findings of the Kepler mission unveiled much of the uncertainty about extra-solar planets (exoplanets), reducing the uncertainty in estimating the abundance of extraterrestrial life. One of the unknown "astronomical" parameters in the Drake Equation has been the fraction of stars with planets, and in particular Earth-like planets. As this work applies the recent exoplanet findings to the abundance of biotic planets and the Drake equation, a few exoplanet key results should be reviewed. In the past decade planets have been discovered around hundreds of nearby stars [Fridlund *et al*. 2010]. However, until the Kepler mission the main method to detect exoplanets has been the Doppler (radial motion) method, which is strongly biased towards massive planets. Of some 400 exoplanets found until 2010 almost all have been Jupiter-sized. The data released by the Kepler mission revealed over 2000 exoplanet candidates, most of which are smaller than Neptune and their planet mass histogram is peaked towards the smaller end of a few Earth masses [Borucki et al 2011a,b; Batalha et al., 2012]. Small planets have been shown to be abundant [Buchhave et al., 2012] and likely found in the habitable zone [Traub, 2011]. Recent analyses of the Kepler statistics showed that about 20% of all Sun-like stars have Earth-sized planets orbiting within the habitable zone [Petigura, Howard and Marcy 2014]. These results are confirmed by observations other than Kepler. The HARPS team estimated that more than 50% of solar-type stars harbor at least one planet, with the mass distribution increasing toward the lower mass end (<15 Earth masses) [Mayor et al., 2011]. Using microlensing observations it has been estimated that on average there is at least one planet per star in the Galaxy [Cassan et al., 2012]. These findings demonstrate that Earth-like planets are probably quite common, enhancing the probability to find planets with conditions appropriate for the evolution of biological life, as we know it. Based on these results we may now better estimate the abundance of life in our stellar neighborhood [Wandel 2011; 2013].

There are two approaches to the search for life outside of the Solar system: (i) looking for biotic signatures (biomarkers) in the spectra of Earth-like extra solar planets, and (ii) searching for intelligent electromagnetic signals (SETI). The success probability of both methods depends critically on the distance to the nearest candidates. The first method requires spectral analyses of extra solar planets, looking for water vapor and gases produced by biotic systems, like oxygen (photosynthesis) or methane. Spectroscopy of exoplanets is marginally possible with present technology, and even with planned future projects it may be feasible only for relatively nearby exoplanets. In addition to the technological challenge, spectral methods could miss many potential life forms; as alien life may be very different from life on Earth, we could fail to recognize it by spectral analyses.

The second approach circumvents these limitations by looking not for the physical signs of biotic life, but rather for technological, radio (or other electromagnetic radiation) broadcasting civilizations. Also this strategy, however, has drawbacks: complex life and in particular intelligent life is presumably much scarcer than simple, mono-cellular life, and hence the distances to the nearest broadcasting civilizations may be very large, possibly beyond the detection range of present, and perhaps even future radio telescopes.

In order to assess the feasibility of detecting biotic planets and intelligent extraterrestrial signals it is essential to estimate the distances to the eventual targets. This work goes in this direction by applying the recent results from the Kepler mission. In order to estimate the abundance of biotic life I derive useful expressions and figures for the distance to biotic exoplanets and to putative civilizations. It is suggested that the probability for the evolution of biotic life may be estimated by future spectral observations of exoplanet biomarkers. Similarly it is argued that future planned radio telescopes may constrain the abundance of radio loud civilizations.

**How far is the closest biotic planet?**

How common are worlds harboring life? The recent findings of Kepler indicate that Earth-sized planets may be found around almost every star. However, assuming that life may develop only on Earth-like planets orbiting Sun-like stars (an assumption likely to be too conservative, as alien life may develop in environments very different from Earth's biosphere), the number of candidates may be reduced to about 10% of all stars.

As is well known, the Drake equation (eq. 5 below) estimates the number civilizations in the Milky Way (also referred to as "the Galaxy"). By analogy, the number of biotic planets in the Galaxy, $N_b$, may be assessed by a "biotic Drake equation"

$$N_b = R^* F_s F_p F_e n_{hz} F_b L_b. \tag{1}$$

The first five terms are astronomical factors and the last two may be called "biotic parameters". The astronomical factors include the rate of star birth in the Galaxy, $R^*$, the fraction of stars suitable for evolution of life, $F_s$, and three "planetary" factors: the fraction of stars that have planets, $F_p$, the fraction of Earth-sized planets, $F_e$, and the number of such planets within the Habitable Zone, $n_{hz}$. These five "astronomical factors" can be combined into a single parameter $R_b$, the rate at which stars suitable for the evolution of biotic life are formed in the Galaxy,

$$R_b = R^* F_s F_p F_e n_{hz}. \tag{2}$$

The average star birth rate in the Galaxy is $R^* \sim 10$ yr$^{-1}$ [e.g. Carroll and Ostile, 2007]. If evolution of life is assumed to be limited to stars similar to our Sun, then $F_s \sim 0.1$. However, this assumption is probably a too restrictive hypothesis. The Kepler data show that Earth size planets are frequent within the Habitable Zone of lower Main Sequence small stars [e.g. Dresing and Charbonneau, 2013], which are the majority of all stars. If life can evolve on planets of red dwarfs [Guinan and Engle, 2013, Scalo, et al., 2007] then $F_s \sim 1$ (since 75% of all stars are red dwarfs).

Recent exoplanet findings, in particular those by the Kepler mission, suggest that probably most stars have planetary systems, hence $F_p \sim 1$. Analyses of the Kepler results shows that 7-15% of the Sun-like stars have an Earth-sized planet within their habitable zone [Petigura et al., 2014], which gives $F_e n_{hz} \sim 0.1$. If biotic life is not restricted to Earth-like planets and to the Habitable Zone (e.g. as in the case of Jupiter's moon Europa) then $F_e n_{hz}$ may be even bigger, up to order unity. Combining all these factors gives for the product of the astronomical parameters in equation (1) a probable range of $0.1 < R_b < 10$ yr$^{-1}$.

The "biotic parameters" in eq. (1) are $F_b$, the probability of the appearance of biotic life within a few billion years on a planet with suitable conditions, and $L_b$, the average longevity of biotic life in units of Gyr ($10^9$ years). Equations (1) and (2) give

$$N_b \sim 10^{11} (R_b/R^*) F_b (L_b/10), \tag{3a}$$

Similarly, the space density of biotic planets, $n_b$, may be written as

$$n_b \sim n^* (R_b/R^*) F_b (L_b/10), \tag{3b}$$

where $n^* \sim 0.01$ ly$^{-3}$ is the average space density of stars in the Galaxy. Considering the history of life on Earth, $L_b$ is likely to be at least a few billion years, so for stars similar to the Sun (or smaller) $L_b \sim 10$ ($10^{10}$ yr). Substituting the values of $L_b$ and $R^*$ equations (3a,b) become

$$N_b \sim 10^{10} R_b F_b, \tag{4a}$$

$$n_b \sim 0.001 R_b F_b \text{ ly}^{-3}. \tag{4b}$$

Since the average distance between biotic worlds is $d_b \sim n_b^{-1/3}$, eq. (4b) gives

$$d_b \sim 10 (R_b F_b)^{-1/3} \text{ ly}. \tag{4c}$$

Eq. (4c) gives the probable distance to our nearest biotic neighbor, plotted in Fig. 1, as a function of the biotic factor $F_b$.

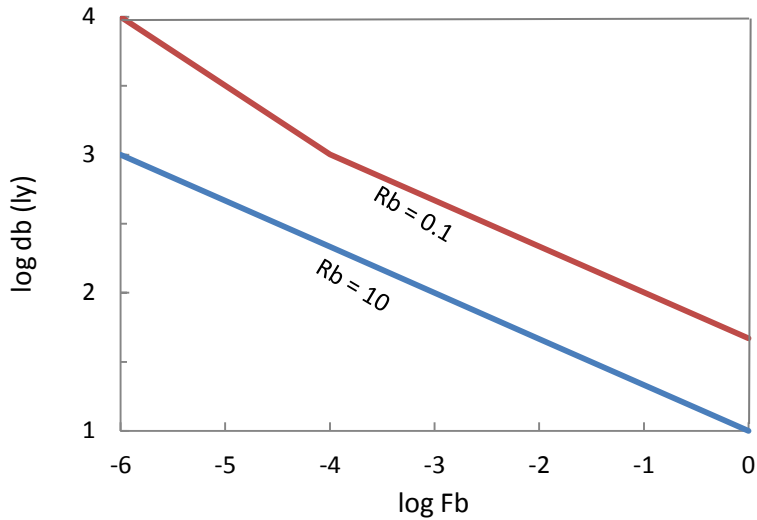

**Figure 1.** The probable distance to our nearest biotic neighbor, $d_b$ vs. the biotic factor $F_b$, for two values of the parameter $R_b$. For the break in the upper curve see the discussion following eq. (7a).

**Estimating the Biotic Parameter**

Given the recent progress in understanding the demographics of planets, in particular Earth-like planets (the "astronomical factor"), the biotic parameter, namely the probability of the evolution of biotic life, remains the major missing factor for estimating the distance to our neighbor biotic planets.

Simple life has appeared on Earth 3.5-3.8 Gyr ago - less than 1 Gyr after Earth's formation and merely ~0.1 Gyr after the establishment of appropriate environmental conditions (the end of the heavy bombardment), a time very short compared to the ages of Earth and the stars. *If* Earth is a typical case (the mediocrity principle), it is plausible that $F_b$ is close to unity. On the other hand, as the origin of life on Earth is very poorly understood, some experts in that area assign an extremely low probability to a similar scenario happening elsewhere (the "Rare Earth Hypothesis"; Benner et al. 2002). According to this approach $F_b$ may be extremely small.

Bayesian analysis demonstrates that as long as Earth remains the only known planet with biotic life, any value could be assigned to $F_b$ [Turner, 2012]. Hence the abundance of extraterrestrial biotic life remains an open question, depending on the (yet unknown) probability for biotic life to evolve on planets with suitable conditions within a given cosmic time. However, a breakthrough in this area may be reached in the near future, if spectral characteristics such as oxygen and other biomarker gases are detected in the atmospheres of Earth-like exoplanets, e.g. by studying eclipses[1] [Rauer et al. 2011; Palle et al. 2011; Loeb and Maoz 2013] or by advanced space telescopes, such as the James Webb Space Telescope[2] (JWST), the Darwin mission[3]

---

http://sci.esa.int/sre-fa/47037-exoplanet-spectroscopy-mission-esm (oct 2014)/ [1]
http://www.jwst.nasa.gov (oct 2014)/ [2]
http://www.esa.int/Our–Activities/Space–Science/Darwin–overview oct 2014)/ [3]

and the Terrestrial Planet Finder[4]. If a few planets with *bone fide* biosignatures are found, the biotic parameter $F_b$ could be assigned an approximate value. Quantitatively, suppose that out of a sample of $N_c$ appropriate candidates (Earth-like planets within the Habitable Zone of a star aged at least a few Gyr), biosignatures are detected in the spectra of $N_{bs}$ planets. Straightforward probability argumentation would imply that the likelihood of biotic life evolution is of the order of $F_b \sim N_{bs}/N_c$. This value should be modified to account for eventual selection effects and sampling corrections. On the other hand, if out of the above sample of $N_c$ candidate planets *no* one shows a definite biosignature, this null result would impose an upper limit $F_b < 1/N_c$. If and when the number of planets with biosignatures grows, a probability distribution function may be constructed, depending on planet lifetime, size, parent star type etc.

Assuming $F_b$ is in the range of $0.001 < R_b F_b < 1$, eq. (4a) gives that the number of biotic planets in the Milky Way is between millions and billions, and the corresponding distance to the nearest biotic world (given by eq. (4c)) is between 10 and 100 light years.

**The distance to putative civilizations**

In this section we extend the above analyses to putative civilizations and apply it to the Search for Extra Terrestrial Intelligence (SETI). The number of civilizations in the Milky Way ($N_c$) may be expressed by the Drake equation,

$$N_c = R^* F_s F_p F_e n_{hz} F_b F_c L_c . \qquad (5)$$

Similarly to eq. (1), the eight variables on the right hand side may be divided into two groups: the first five are astronomical factors and the last three - one biotic ($F_b$) and two "civilization" factors. The latter two are related to the development of intelligent life: the intelligence-communicative factor $F_c$, defined as the probability that one or more of the species on a planet harboring biological life will eventually develop a civilization using radio communication, and $L_c$, the broadcasting longevity (in years) of such a civilization.

Expressions analogous to eqs. (4a,b), for the number of communicative civilizations ($N_c$) and their abundance (space density, $n_c$) are easily derived by replacing $L_b$ in eqs. (3a,b) with $F_c L_c$,

$$N_c \sim 1000\, R_b F_b F_c L_{c3} \qquad (6a)$$

and

$$n_c \sim 10^{-10}\, R_b F_b F_c L_{c3}\, \text{ly}^{-3}. \qquad (6b)$$

where $L_{c3} = L_c/(1000 \text{ years})$. The normalization $L_{c3}$ is just for convenience and does not presume any specific value to the longevity of radio-communicative civilizations (this will be discussed later).

Even if communicative civilizations thrive in the Milky Way, we may be unable to detect their radio signals, unless the typical distance between them is within our detection range. As the parameter $R_b$ is relatively well estimated, and $F_b$ may be estimated in the near future, e.g. by finding biosignatures as discussed above, the major uncertainty in eqs. (6a,b) remains in the two unknown "civilization parameters", the probability for the evolution of civilizations using radio communication, $F_c$, and their longevity (the duration of their "radio loud" phase) $L_c$. By analogy to eq. (4c) the average distance between communicative civilizations can be derived from eq. (6b),

$$d_c \sim 2000\, (R_b\, F_b\, F_c\, L_{c3})^{-1/3}\ \text{ly} \qquad \text{for } d_c < 1000\ \text{ly} \qquad (7a)$$

Eq. (7a) assumes that the average distance between neighbor civilizations is smaller than the smallest dimension of the system. Since the Galaxy has a shape of a thin disk, this assumption is valid only when the distance $d_c$ is smaller than the scale height of the stellar distribution in the Galaxy (the thickness of the disk), which is ~1000 ly (in other words, as long as the product $R_b\, F_b\, F_c\, L_{c3} > 10$). Otherwise (if $R_b\, F_b\, F_c\, L_{c3} < 10$), we must take into account the flat geometry of the Galaxy. In that case we may calculate the average distance by assuming that $N_c$ civilizations are uniformly distributed on the area of the Galactic disk (a circle with a diameter of $d_G \sim 100{,}000$ light years). Equating this to the area of $N_c$ circles each having a radius $d_c$ gives $\pi d_c^2 N_c \sim \pi d_G^2$ and the average distance between neighbor civilizations is $d_c \sim 10^5\, N_c^{-1/2}$ ly. Substituting $N_c$ from eq. (6a) gives for this case

$$d_c \sim 3000\, (R_b\, F_b\, F_c\, L_{c3})^{-1/2}\ \text{ly} \qquad \text{for } d_c > 1000\ \text{ly} \qquad (7b)$$

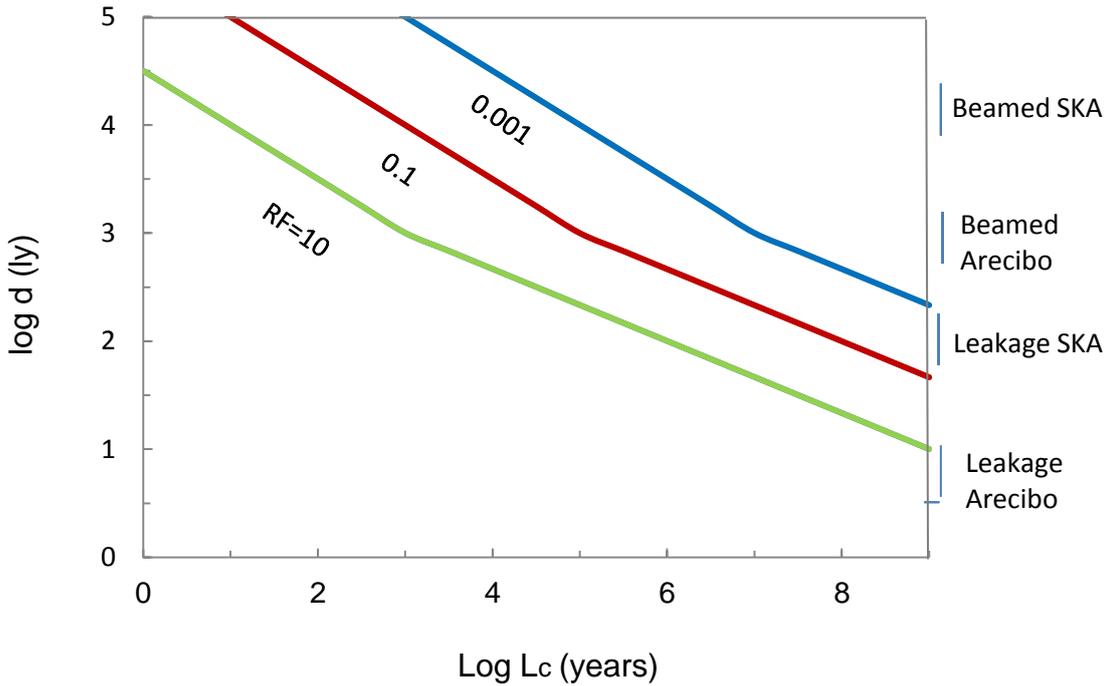

**Figure 2**. The average distance between neighbor civilizations *(d)* vs. the average longevity of a communicative civilization $L_c$, for several values of the product $RF = R_b\, F_b\, F_c$. On the vertical *right* axis are marked the relevant detection ranges of leakage and beamed radio signals by the Arecibo and SKA telescopes (see text).

Fig. 2 shows the average distance between neighbor civilizations (eqs. 7a,b) vs. the average longevity of a communicative civilization $L_c$, for several values of the product $RF = R_b\, F_b\, F_c$.

Unlike the biotic factor, the civilization parameters $F_c$ and $L_c$ cannot be inferred even intuitively from the evolution of life on Earth. As the evolution of complex life on Earth took about ~ 4 billion years, the probability for the evolution of complex and intelligent life during a few billion years could be anywhere between unity (corresponding to Earth being a typical case) and very small (if the appearance of complex life and intelligence is an extremely rare event [e.g. Cuntz et al. 2012] or typically requires a much longer time. Also $L_c$, the average longevity of a communicative civilization, cannot be inducted from its short history on Earth and could be anywhere between a few hundred years and billions of years. Communicative civilizations may disappear within a relatively short time after developing radio technology because of self destruction (wars or ecological disasters) or else become less detectable due to the development of radio-quiet communication channels. On the other hand, civilizations may survive such "childhood diseases" by spreading to other planets and last much longer. Obviously the weighted average of the longevity would be increased by older, long lasting (and broadcasting) civilizations.

**The SETI success probability**

There are two scenarios for detecting radio signals from extraterrestrial civilizations: (1) finding a purposeful, directed broadcast attempt, including an interstellar automatic radio beacon or (2) civilizations may be detected through no special efforts of their own. The latter hypothesis, often called eavesdropping, is concerned with the extent to which a civilization can be unknowingly detected through the by-products of its daily activities, e.g. the leakage of its own radio communication to space. The range for detecting such radio signals depends on the receiver sensitivity and on the transmitting power, as well as on the level of the background noise and on whether the signal is directed or isotropic. Beamed transmissions directed at a specific target would be much stronger and thus detectable from longer distances than the semi-isotropic broadcasting, typical for radio stations, looked after by eavesdropping. Future radio observatories such as EVLA (Expanded Very Large Array), LOFAR (Low Frequency Array) and Square Kilometer Array (SKA) may be able to detect low-frequency radio broadcast leakage from a civilization with a radio power similar to ours out to a distance of a few hundred of light years [Loeb and Zaldarriaga, 2007]. SKA would be able to detect an airport tower radar from 30 light years[5]. Beamed transmissions could be detected over much larger distances. For example, a targeted search by the Arecibo telescope could detect alien signals sent by a similar device (i.e., with a similar power, ~$10^{13}$ Watt/m$^2$/radian$^2$) and aimed at Earth from distances of a few thousands of light years. Noteworthy, the sensitivity of all-sky surveys is much lower and the above detection ranges need to be decreased by a factor of 10 -- 100. Eqs. (7a,b) imply that even with rather "optimistic" values of the other parameters ($R_b\, F_b\, F_c \sim 1$), unless $L_c$ is longer than a million years, the average distance between neighbor civilizations is thousands of light years, far beyond the range of eavesdropping even by future telescopes such as SKA.

---

[5] /https://www.skatelescope.org/key-documents (oct. 2014)

Transmissions beamed at Earth, either unintentionally, like an aviation radar or communication satellite, or intentionally like the Arecibo message of 1974[6], may be detected from considerably larger distances. For example, a beamed transmission at the broadcasting power similar to that of the Arecibo radar can be detected by the Arecibo radio telescope at a range of ~ 3000 light years, and by future telescopes such as SKA the detection range may increase to ~30,000 light years, virtually across the Milky Way. These ranges are shown on the right vertical axis in fig. 2. Note however that the effective broadcasting time of beamed signals may be significantly shorter than the total communicative longevity, as discussed below.

The expressions for the average distance between civilizations derived above can be used to estimate the success chances of SETI by calculating the probability that a broadcasting civilization happens to lie within the detection range of present and future radio telescopes. Let us first consider eavesdropping (looking for leakage signals). Using the sensitivity of future radio telescopes such as SKA, the eavesdropping detection range is of the order of 100 light years. Equation (7a) shows that a neighbor civilization is likely to exist within this distance if the "Drake product" $F_b F_c L_c$ is of the order of a million years or more. For smaller values of the Drake product we may define the likelihood $p(d)$, that a civilization happens to exist at a distance $d < d_c$, that is, closer to Earth than the average distance between neighboring civilizations. A straightforward geometric approach (fractional volume) gives

$$p(d) \sim (d/d_c)^3. \qquad (8)$$

Fig. 3 shows the likelihood $p(d<100)$, that a broadcasting civilization happens to exist within a distance of 100 ly from Earth, as a function of the average longevity of communicative civilizations, for several values of the Drake product.

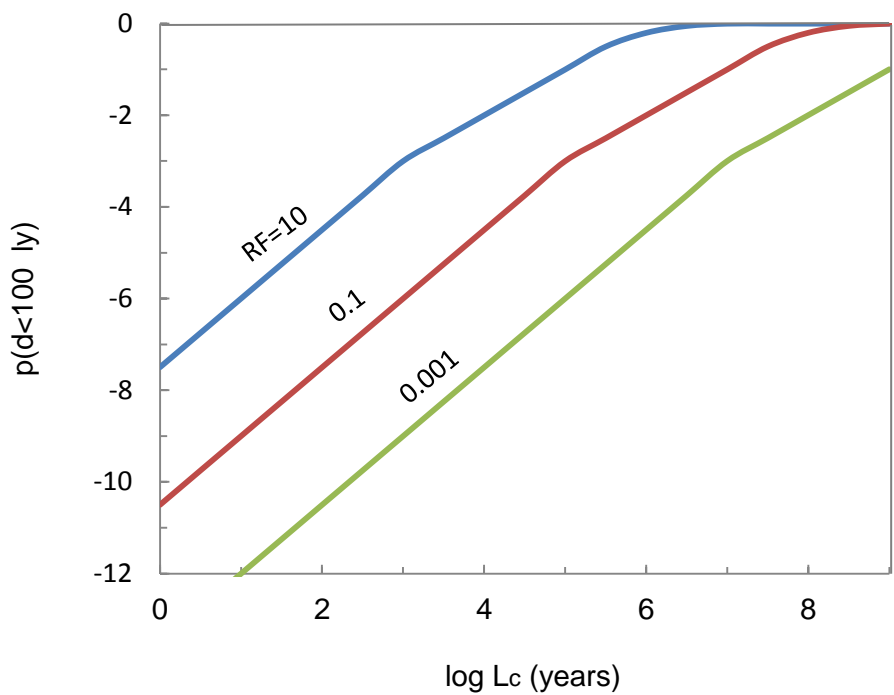

---

**Figure 3**. The probability of a civilization to exist within a distance of 100 light years from Earth, vs. the average longevity of a communicative civilization, $L_c$, for several values of the product $RF = R_b F_b F_c$.

Combining eqs. (8) and (7a,b) gives the probability to detect leakage signals in terms of the Drake parameters (assuming a detection range of 100 ly, appropriate for planned radio arrays such as SKA):

$$p(d < 100 \text{ ly}) \sim 10^{-4} R_b F_b F_c L_{c3} \qquad d_c < 1000 \text{ ly}, \qquad (9a)$$

$$p(d < 100 \text{ ly}) \sim 3 \cdot 10^{-5} (R_b F_b F_c L_{c3})^{3/2} \qquad d_c > 1000 \text{ ly}. \qquad (9b)$$

As in eq. (7a), the condition on $d_c$ in eq. (9a) is equivalent to a condition on the product $R_b F_b F_c L_{c3} > 10$, and in eq. (9b) $R_b F_b F_c L_{c3} < 10$. For example, if we assume that communicative civilizations are common ($R_b F_b F_c \sim 1$), and if they typically transmit for ten thousand years ($L_{c3} \sim 10$), the probability that a civilization is presently broadcasting within ~ 100 ly from Earth is 0.001.

**The probability of detecting beamed signals**

As discussed above, assuming the broadcasting power of putative civilizations is comparable to ours, eavesdropping on leakage signals has a quite limited range; the Arecibo telescope could detect leakage signals from civilizations broadcasting at the power of Earth at a distance of only a few light years, and future telescopes such as SKA up to ~100 ly. On the other hand, beamed transmissions may be detected from considerably larger distances. For example, a beamed transmission comparable to that of the Arecibo radar may be detected by the Arecibo dish at a range of ~ 3000 ly, and by future telescopes such as SKA up to ~30,000 ly. Apparently, these increased detection ranges should give a significantly higher detection probability. On the other hand, beamed signals may not be continuously aimed in our direction, contrary to leakage signals. This effect is likely to reduce the effective broadcasting duration. It may be described by introducing a "beaming parameter" $b$, the fraction of the broadcasting lifetime of a civilization during which it is actually sending signals beamed in our direction. This beaming may happen either unintentionally, as communication with satellites, spacecrafts, or planets in their own solar system, or deliberately as interstellar messages [e.g. Zetkov 2006]. In other words, $bL_c$ is the integrated duration of beamed broadcasting in the direction of Earth.

For $R_b F_b F_c b L_{c3} < 10$ (see eq. 9b) the probability to find a civilization within the detection range of beamed signals by Arecibo end SKA, respectively, is

$$p(d < 3000 \text{ ly}) \sim 1 (R_b F_b F_c b L_{c3})^{3/2} \qquad \text{Arecibo}, \qquad (10a)$$

and

$$p(d < 30{,}000 \text{ ly}) \sim 1000 (R_b F_b F_c b L_{c3})^{3/2} \qquad \text{SKA}. \qquad (10b)$$

For example, assuming that communicative civilizations are common, that is $R_b F_b F_c \sim 1$, and that on average signals are beamed at Earth during an integrated time of $bL_c \sim 10$ years, equations (10a,b) give a detection probability $p \sim 10^{-3}$ by Arecibo and $p \sim 1$ by SKA. If actually $R_b F_b F_c b L_{c3} < 10$ it is not surprising that SETI has not yet detected an alien signal (the "Great Silence"), and it may remain silent even with the increased sensitivity of future telescope arrays. If, on the other hand, $R_b \, F_b \, F_c \, b \, L_{c3}$ >10, then applying eq. (9a) to beamed signals gives

$$p(d< 3000 \text{ ly}) \sim 10^{-4} R_b \, F_b \, F_c \, b \, L_{c3} \qquad \text{Arecibo},$$

and

$$p(d<30{,}000 \text{ ly}) \sim 1 \qquad \text{SKA},$$

implying that SKA might be able to detect signals beamed at Earth by putative civilizations.

**Estimating the Civilization Parameters**

Similarly to the biotic factor, also the civilization parameters $F_c$ and $L_c$ are presently unknown. If SKA and advanced future radio arrays fail to detect intelligent extraterrestrial signals, eqs. (9a,b) and (10b) may be used to place an upper limit on the products $F_c \, L_c$ and $F_c \, bL_c$, respectively. On the other hand, if an extraterrestrial signal from another civilization is detected, the present ignorance in the civilization parameters may be removed, or at least significantly constrained. If by then also biomarkers of exoplanets are detected, so that $F_b$ can be estimated to some extent, the product $F_c \, L_c$ could be assigned an approximate value; suppose that out of a sample of $N_{bs}$ suitable civilization candidates (that is, Earth-like planets aged at least a few Gyr with a biosignature) intelligent signals are detected from $N_{is}$ ones; here, in analogy to the section on the biotic parameter above, $N_{bs}$ is the number of planets with biotic signature, and $N_{is}$ is the number of planets from which intelligent signals are detected. A similar argument as in the case of the biotic factor would count for the "communicative factor", or the likelihood of a communicative civilization to evolve on a biotic planet, multiplied by the average broadcasting longevity, leading to a value of the order of

$$F_c \, L_c \sim 10 \text{ Gyr } N_{is} / N_{bs},$$

where 10 Gyr is the age of the Galaxy. As in the case of the biotic fator, this expression needs to be modified by eventual selection effects and sampling corrections, as well as by putative contributions to $N_{is}$ from non biotic planets such as automated beacons.

**Summary**

The recent results of the Kepler mission significantly reduce the uncertainty in the astronomical parameters of the Drake equation. I derive expressions for the space density of biotic worlds as a function of the (yet unknown) probability for the evolution of biotic life and the uncertainty in the astronomical parameters. Similar expressions are derived for the distance to putative communicative civilizations, depending on two additional unknown factors in the Drake equation, the probability of evolution from simple biotic life to a communicative civilization and its longevity. Additionally, the probability to detect radio signals from other civilizations with present and future radio telescopes is estimated in terms of these factors. The extended analyses, updated by the Kepler results, presented in this paper suggests that our nearest biotic neighbor exoplanets may be as close as 10 light years. Even with a less optimistic estimate of the biotic probability, for example that biotic life evolves on one in a thousand suitable planets, our biotic neighbor planets may be expected within 100 light years. On the other hand, the distance to the nearest putative civilizations, even for optimistic values of the Drake parameters, is estimated to be thousands of light years.

## References

Batalha, N. M., Rowe, J. F., Bryson, S. T., et al. 2012, Planetary candidates observed by Kepler, iii: analysis of the first 16 months of data, Astrophysical Journal Supp., 204, 24B

Benner, S. A., Caraco, M. D., Thomson, J. M., Gaucher , E. A. May 2002, Planetary Biology--Paleontological, Geological, and Molecular Histories of Life, Science 296 (5569): 864–868

Borucki, W. J., Koch, D. G., Basri, G., et al. 2011b, Characteristics of planetary candidates observed by Kepler, ii: analysis of the first four months of data, Astrophysical Journal, 736, 19

Borucki, W. J., Koch, D. G., Basri, G., et al. 2011a, Characteristics of Kepler planetary candidates based on the first data set, Astrophysical Journal, 728, 117

Bounama, C., von Bloh, W. and Franck, S. 2007, Astrobiology, 7(5): 745-756

Buchhave, L. A., Latham, D. W., Johansen, A., et al. 2012, An abundance of small exoplanets around stars with a wide range of metallicities, Nature, 486, 375

Cassan, A; Kubas, D.; Beaulieu, J.-P.; et al. 2012, One or more bound planets per Milky Way star from microlensing observations, Nature 481 (7380): 167–169

Carroll, B.W and Ostile, D.A. 2007, An Introduction to Modern Astrophysics, Pearson p. 1019

Cuntz, M., von Bloh W., Schroeder, K-P. et al. 2012, Habitability of super-Earth planets around main-sequence stars including red giant branch evolution: models based on the integrated system approach, International Journal of Astrobiology, 11,15-23

Dressing, C.D. and Charbonneau, D. 2013, The occurrence rate of small planets around small stars, Astrophysical Journal 767, 95

Fridlund M., Eiroa, C., Henning, T. et al. 2010, The search for worlds like our own, Astrobiology 10:5-17

Guinan, E. F. and Engle, S. G. 2013, Assessing the suitability of nearby Red Dwarf stars as Hosts to Habitable Life-bearing, Proc of the American Astronom. Soc. 221, 333.02

Loeb, A. and Maoz, D. 2013, Detecting bio-markers in habitable-zone earths transiting white dwarfs, Mobthly Notices Royal Astronom. Soc., 432, L11

Loeb A. and Zaldarriaga, M. 2007, Eavesdropping on Radio Broadcasts from Galactic Civilizations with Upcoming Observatories for Redshifted 21cm, Radiation, J. Cosmology and Astroparticle Phys. Jan. (Issue 1: article #20)

Mayor, M. , Marmier, M. , Lovis, C. et al. 2011, The HARPS search for southern extra-solar planets XXXIV. Occurrence, mass distribution and orbital properties of super-Earths and Neptune-mass planets, Astronomy. & Astrophys, 541, 139

Rauer, H., Gebauer, S., Paris, P. V., et al. 2011, Potential biosignatures in super-Earth atmospheres. I. Spectral appearance of super-Earths around M dwarfs, Astronomy. & Astrophys, 529, A8

Palle, E., Zapatero Osorio, M. R., and García Munoz, A. 2011, Characterizing the atmospheres of transiting rocky planets around late-type dwarfs, Astrophysical Journal, 728, 19

Petigura, E. A., Howard, A. W. and Marcy, G. W. 2014, Prevalence of Earth-size planets orbiting Sun-like stars, PNAS 2013 110 (48) 19175-19176, arXiv:1311.6806

Scalo, J., Kaltenegger, L., Segura, A. et al. 2007, M Stars as Targets for Terrestrial Exoplanet Searches And Biosignature Detection Astrobiology, 7(1): 85-166.

Strobel, D. F. 2010, Molecular hydrogen in Titan's atmosphere: Implications of the measured tropospheric and thermospheric mole fractions. *Icarus*, DOI: 10.1016/j.icarus.2010.03.003

Tarter J. 2001, The Search for Extraterrestrial Intelligence (SETI), Ann. Rev. Astr. Astrophys. 39: 511-48

Traub, W. A.2011, Terrestrial, habitable-zone exoplanet frequency from kepler, Astrophysical Journal, 745, 20

Turner, E.L. 2012, unpublished talk at the 293 IAU symposium on Extrasolar Habitable Planets

Wandel, A. 2011, The impact of Kepler on the chances of extraterrestrial life, proc. of the annual meeting of ILASOL, http://www.ilasol.org.il/uploads/files/ILASOL_25th-Abstracts-271111.pdf

Wandel, A. 2013, How frequent is biotic life in space? proc. of the annual meeting of ILASOL, http://www.ilasol.org.il/uploads/files/Wandel2013.pdf

Zaitsev, A. 2006,, The SETI Paradox, Bull. Spec. Astrophys. Obs., 60